\documentclass[english]{article}
\usepackage{hyperref}
\usepackage[utf8]{inputenc}
\usepackage[T1]{fontenc}
\usepackage{babel}

\usepackage{mathtools}
\usepackage{graphicx}
\usepackage{fancyhdr}
\usepackage{url}
\usepackage{siunitx}

\usepackage{lscape}
\usepackage{nicefrac}
\usepackage[superscript]{cite}

\usepackage{color}
\definecolor{blue}{rgb}{0,0,1}
\newcommand{\revise}[1]{{#1}}

\usepackage{listings}
\begin{document}

\title{\revise{Database of Power Grid Frequency Measurements}}

\author{Richard Jumar\textsuperscript{1{*}}, 
Heiko Maa\ss\textsuperscript{1}, 
Benjamin Sch\"afer\textsuperscript{2},\\
Leonardo Rydin Gorj\~ao\textsuperscript{3,4}, and
Veit Hagenmeyer\textsuperscript{1}
}

\maketitle
\thispagestyle{fancy}
\noindent 1. Karlsruhe Institute of Technology - Institute for Automation and Applied Informatics (IAI), Germany \\
2. School of Mathematical Sciences, Queen Mary University of London, United Kingdom \\
3. Forschungszentrum J\"ulich, Institute for Energy and Climate Research - Systems Analysis and Technology Evaluation (IEK-STE), Germany\\
4. Institute for Theoretical Physics, University of Cologne, Germany\\
{*} corresponding author: Richard Jumar (richard.jumar@kit.edu)
\begin{abstract}
\revise{The transformation of the electrical energy system attracts much attention in diverse research communities. Novel approaches for modeling, control, and power grid architectures are widely proposed. Data from actual power system operation are therefore critical to evaluate models and to analyze real-world scenarios. However, they are rarely available. In the present paper, we introduce a precisely time-stamped data set of power grid frequency measurements. It covers twelve synchronous areas of different sizes in one-second resolution. The contribution includes synchronized measurements within the Continental European synchronous area acquired in Portugal, Germany, and Turkey, maximizing the geographical span. Finally, we provide excerpts of the underlying waveform. Data were collected using a self-developed measurement instrument, the Electrical Data Recorder (EDR), connected mostly to conventional power sockets. We close our description with a discussion on measurement error and data quality.
}  
\end{abstract}

\section{Background \& Summary}
\begin{figure*}[t!]
\includegraphics[width=0.99\linewidth]{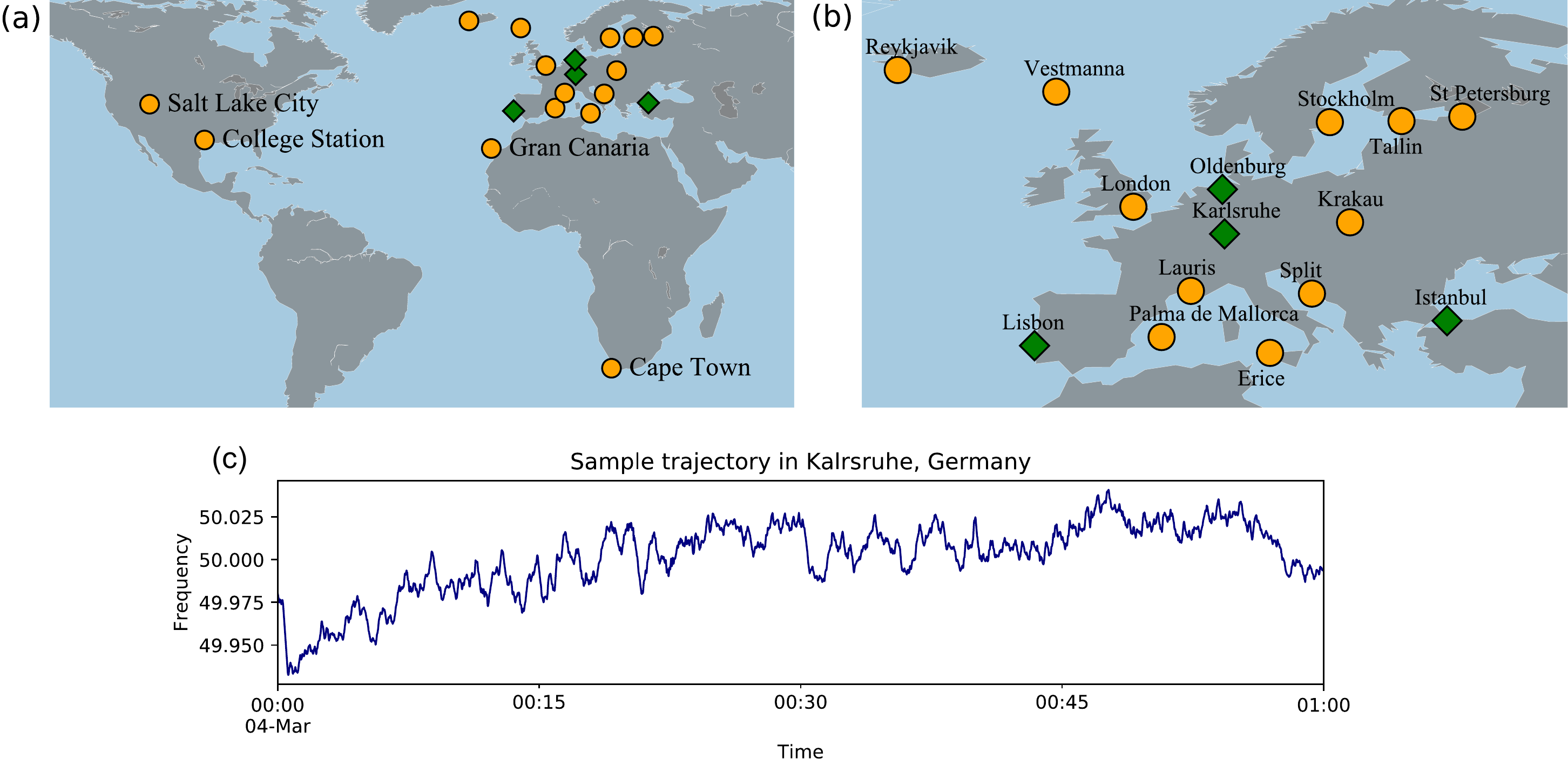}
\caption{Overview \revise{and illustration} of available data. (a): A (partial) world map shows the different locations at which measurements were taken. 
Australia and large parts of Asia are not shown as there were no measurements recorded.
(b): Zoom of the European region (not showing Gran Canaria). 
Circles indicate measurement sites where single measurements for several days were taken, diamonds mark the four locations where we performed synchronized measurements in the Continental European system for one month.
(c): Sample frequency trajectory, recorded in Karlsruhe, Germany.
Maps were created using Python 3 and geoplots.}
\label{fig:Map}
\end{figure*}
For the actual transformation of the energy system, energy research should be transparent to enable scientific discussion, but also particularly for providing information to policymakers or the general public \cite{pfenninger2017energy}. 
In this context, a real-world power grid frequency open data base can be valuable to understand state-of-the-art power systems but also to help to advance the transition towards fully renewable power systems. 

\revise{
Power grid frequency is particularly interesting as it is used as a control signal for the dispatch of generation under free governor action.
The amount of generation from controllable power plants is individually governed by a control curve that relates frequency and power output.
In an interconnected system, multiple generators can therefore share the total grid load based on local measurements of the frequency.
Hence, the frequency fluctuates by design and is an indicator for grid-wide balance of supply and demand.
It is used to assess the system's stability \cite{Machowski2011} and it allows insights in behavior of loads and (also renewable) generators \cite{Schaefer2017a}. 
Frequency dynamics are regulated by the transmission system operators (TSOs) \cite{ENTSOE2013} in a multi-stage approach in the effort to accommodate the needs of energy markets and system stability alike.
Not surprisingly, frequency dynamics were found to be influenced by market activities \cite{weissbach2009}.}

\revise{
In conventionally supplied power grids, mainly the load varies over time, while the generation adapts to meet the demand. 
With the increasing share of less-controllable, renewable sources like photovoltaics and wind, new sources of fluctuations are introduced.
The increasing share of generation from wind has already been found to correlate strongly with the amplitude of frequency fluctuations \cite{Adeen2019}.
Along with this transition, the number of synchronous generators is reduced, which in turn reduces the system's inertia. 
This poses potential challenges for the grid operators, who have to ensure the stability of supply\cite{LLNL-Freq2010}.
Smaller synchronous regions with high penetration of renewable sources are especially critical.
For example, the Irish TSOs, EirGrid, and SONI have analyzed the impact of this development on the frequency response of their grid\cite{Doherty2010,Uijlings2013,PPAEnergyltd2013} and adjusted the grid code accordingly\cite{EirGrid2019}.
This code demands that non-synchronous generators also participate in frequency control by varying their active power output in response to the grid frequency.
While in this case an adjusted grid code enables the integration of a large share of non-synchronous renewable generation, general questions about the portability of the approach to other, different grids and especially fully sustainable power systems remain. 
}

\revise{Moving forward, we should thus understand current power systems better.}
Specifically, to which extent specific regulations, energy mixes, consumer fluctuations, etc. impact the frequency dynamics, and thereby grid stability. 
This understanding is crucial to implement new smart grid concepts \cite{Fang2012}, such as modern demand response concepts \cite{Schaefer2015}, the introduction of \emph{flexumers} \cite{barwaldt2018energy}, or to make theoretical predictions, such as: How do fluctuations impact the grid's stability \cite{schafer2017escape,hindes2019network}? 
Specifically, how far does the size of a synchronous area \textemdash like in island grids or potentially small microgrids \cite{Lasseter2004,Kroposki2008} \textemdash has an impact on the frequency stability issues and control efforts to care for?
How do fluctuations spread within one synchronous area \cite{zhang2019fluctuation,HaehnePropagation2019}? 
When is a power grid susceptible to cascading failures and how do these propagate \cite{Simonsen2008,yang2017small,schafer2019dynamical,nesti2018emergent}? 
\revise{
While simulation-only studies can provide answers in some cases, real data are still essential to validate the underlying assumptions.
Some of the research questions have already been touched upon by previous contributions using wide-area synchronized measurements \cite{vanfretti2013spectral, tuttelberg2018estimation,chai_wide-area_2016} and highlighting their scientific value.
Despite the importance, these data are not yet easily accessible and openly available.}
Even more so, data from differently sized synchronous regions to validate scaling laws and investigate the impact of microgrids have, to our knowledge, not been provided openly at all.
\revise{
Typically, only TSOs of large synchronous areas disclose such data publicly.
For instance, we find consistent data over longer periods for Continental Europe\cite{Transnet}, Great Britain\cite{NationalGrid2019}, and the Nordic Grid\cite{fingrid}.
}
\revise{Meanwhile, synchronized recordings from within one grid and data from smaller areas are essentially unavailable.}
There are however, academic projects and spin-offs that have set up Phasor Measurement Unit acquisition networks from which the frequency could be derived.
Well known are FNET/GridEye\cite{PITL2014}, Grid Radar\cite{Gridradar}, and mainsfrequency.com\cite{Netzfrequenz}.
While theses projects provide freely accessible dash-boards with live data views, the underlying databases are not openly available which limits the scientific usability and value.

Therefore, in the present paper, we introduce an open data base of measurements of the power grid frequency recorded in the period 2017 to 2020 in various European and some non-European synchronous areas. 
Measurement locations are highlighted on a map in Figure~\ref{fig:Map}. 
The focus lies on covering multiple synchronous areas of different size, ranging from small regions, such as the Faroe Islands with approximately fifty thousand inhabitants, to the Continental European synchronous area with about half a billion inhabitants. 
Furthermore, we conducted a synchronized measurement at four different locations within the continental European synchronous area for a full month.
Overall, we offer 428 days of recorded frequency trajectories in a generally applicable CSV format.
The database also contains excerpts of the \SI{25}{\kilo\hertz} sampling of the sinusoidal voltage signal recorded at the given connection point.

Accompanying this publication, we present an initial data analysis that focuses on the scaling and spatio-temporal properties of power grid frequency measurements\cite{gorjo2020open}. 

\section{Methods}
\begin{figure*}[t]
\includegraphics[width=0.99\linewidth]{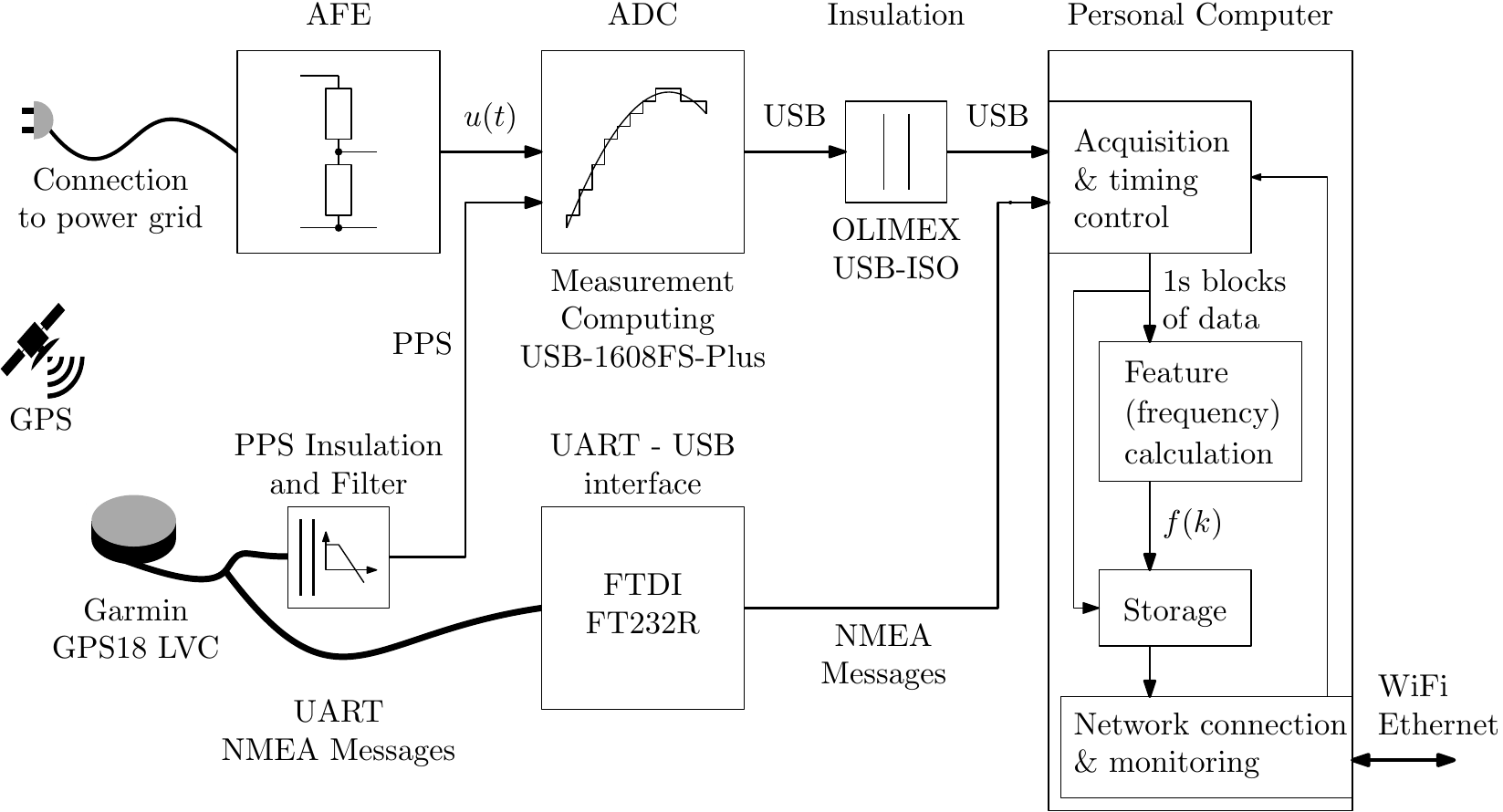}
\caption{Structural overview of the Electrical Data Recorder (EDR).} 
\label{fig:edrSystemOveriew}
\end{figure*}

In the following we describe the details of the data acquisition as well as the signal processing used to generate the time series provided in this data set.
\revise{
We perform data acquisition using an in-house developed instrument, the Electrical Data Recorder (EDR)\cite{EDR_TIM2013, maass2015data, EDR_Rework2019}.
The EDR is connected mostly to conventional power sockets in office or hotel environments to acquire the voltage waveform.
From this signal the frequency is calculated once per second.
In these environments, usually distortions and noise heavily superimpose the voltage signal. 
However, the imperfections of the input only affect the calculated frequency on a sub-millihertz level.
Therefore, the obtained readings depict the grid-wide picture reasonably well. 
We provide more details on this in Section~\ref{sec:TechnicalVal}.}

\revise{
In terms of providing frequency time series with a resolution of one second, the EDR and established synchronized monitoring devices such as Phasor Measurement Units (PMU) are similar.
The advantage of the EDR in this application lies in its portability and ease of use.
However, these two features were key enablers for the conduction of many of the provided measurements. }

\revise{
For a reproduction of the experiment two aspects are the most important: 
(1) the synchronization of the data acquisition and 
(2) the frequency calculation algorithm.
In the following, we describe how the EDR addresses both aspects. 
Thereby, we will see that the device-level description of the EDR is rather straight forward since the instrument relies on established principles.
This means that a reproduction of the experiment does not necessarily require an EDR. 
Any data acquisition system that can be synchronized to an external clock together with the provided frequency calculation algorithm is sufficient. 
Nevertheless, we will close the Section with an in-depth technical description of how the EDR is built and operated.
}

\subsection{\revise{Capturing device: The Electrical Data Recorder}}
\revise{
The EDR is a versatile research-prototype acquisition device to capture waveform-data from the power grid in a synchronized manner.
Figure \ref{fig:edrSystemOveriew} depicts the system architecture. 
\revise{
It comprises a simple analog front end (AFE) for signal conditioning, a USB-connected Analog-Digital-Converter (ADC), a Global Positioning System (GPS) receiver as reference clock, and a Personal Computer.
Additionally, a USB insulation unit and an interface adapter circuit with included PPS filtering are required, which we elaborate on in Section \ref{sec:EDR_Hardware}.
The PC controls the acquisition process and performs the data processing using a self-developed application we call \texttt{EDR\_Scope\_N}.
The EDR is a portable device that allows local data storage as well as online data transmission and remote monitoring when connected to the internet.
The device reports and stores time-stamped aggregated measurements (like the power grid frequency) at one-second resolution.}
Additionally, it is capable of storing the full time series of samples of the underlying electrical signal.
Using this feature is costly in storage requirements but allows the greatest flexibility towards \revise{offline, post-processing} exploitation of the data.
The acquisition time is referenced to the Coordinated Universal Time (UTC), using the timing information provided by the satellite-based GPS and its high resolution Pulse Per Second (PPS) signal.
}

\begin{figure*}[t]
\includegraphics[width=0.99\linewidth]{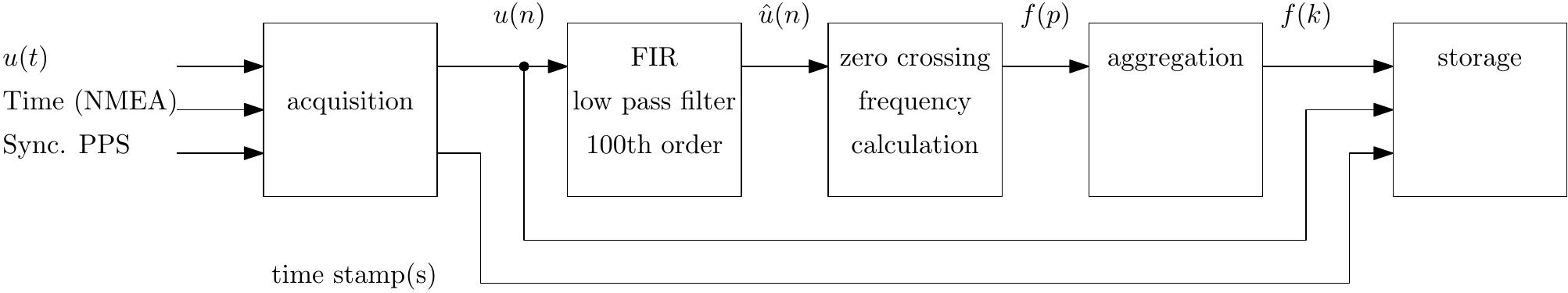}
\caption{Logic representation of the data flow: From continuous voltage input $u(t)$ to one-second aggregates of the power grid frequency $f(k)$.} 
\label{fig:dataflow}
\end{figure*}

\subsection{Acquisition and synchronization principle}
In this section we describe the data acquisition procedure from the physical input signal to the discrete samples.
We acquire the voltage at the connection point, \revise{
usually a conventional wall socket in an office or hotel environment.}

\revise{
The voltage signal (continuous in time $t$) is subsequently scaled} and continuously sampled using a $16$~bit ADC at a nominal sampling frequency $f_s$ of \SI{25}{\kilo\hertz}\footnote{
The sampling rate is adjustable. 
In a few cases we set the rate to \SI{12.8}{\kilo\hertz} to save disc space and to make longer recordings.}.
This process results in a series of discrete sampled values $u(n)$ spaced $\nicefrac{1}{f_s}$ on the time axis.
Since $f_s$ is provided by a quartz oscillator within the ADC it is accurate to approximately $\pm 100\,\text{ppm}$. 

More accurate timing information is provided by an accompanying GPS receiver in two different ways:
Second level timestamps are derived from the standardized NMEA 0183 messages, sent by the receiver once per second as string of ASCII characters. 
The temporal alignment of each sample $u(n)$ is provided by evaluating the receiver's PPS signal.
We feed the low-pass filtered PPS and $u(t)$ to synchronously captured ADC channels.
\revise{
Therefore, the stream of samples of the filtered PPS and $u(n)$ are aligned with each other.
The acquisition software then searches for the PPS transitions in the streams and aligns the whole time series to the PPS reference.
Figure \ref{fig:dataflow} depicts a logical representation of the signal and data flows within the instrument. 
Processing of the samples is performed in blocks of a duration of one second.}
Timestamps are assigned to these blocks by the software based on the slope (rising edge) of the PPS \cite{EDR_Rework2019}.
This results in secondly stamped frequency data, which are subsequently buffered on local hard drive and sent to a large-scale storage when an internet connection is available.
Sampling, evaluation, and storage are designed for continuous long-term time-stamped full-sample data recording without interruptions.

\subsection{Frequency calculation}
\revise{Many different frequency calculation algorithms are known that are all tailored to address specific requirements of certain applications.
PMUs typically apply methods based on the derivation of the instantaneous phase since this a byproduct of the phasor estimation. 
We instead use the Zero-Crossing method, which is probably the most straight forward one.
Although very simple and intuitive, it provides modest robustness against phase and amplitude steps and is (as opposed methods using Discrete Fourier Transform [DFT]) inherently independent of the nominal frequency.
It is popular in embedded devices and also used for power grid frequency measurements\cite{Netzfrequenz}.
Since the method only evaluates a very small section of the signal around the zero crossing (ZC), it is slightly more susceptible to noise than DFT-based approaches.
However, its major drawback is the dependency between input signal frequency and reporting rate.
Since the algorithm only evaluates full cycles of the input waveform, the Zero-Crossing method is limited to rather low reporting rates.
For the content of the present paper,  this is not a restriction since we use a one-second resolution.}

Following the processing chain (Figure~\ref{fig:dataflow}) the digital signal is conditioned prior to the frequency calculation.
Signal conditioning aims at removing distortions (harmonics, injections from photovoltaic inverters, voltage transients, etc.) from the wave form.
In frequency domain this is equivalent to limiting the signals bandwidth to include only the interesting frequency deviations. 
The samples $u(n)$ are therefore low-pass filtered using a Finite Impulse Response (FIR) filter.
Exploiting the linear phase response of this filter class, the constant group delay is counteracted by shifting the sample indices accordingly. 
Hence, there is no resulting delay between the input $u(n)$ and the filtered output $\Tilde{u}(n)$. 
The filter length $L$ and coefficients $h(n)$ are selected based on the nominal power grid frequency $f_\text{nom}$ which can be \SI{50}{\hertz} or \SI{60}{\hertz} and the sampling rate $f_\text{s}$.
Choosing $L = \nicefrac{f_\text{s}}{5 f_\text{0}} \cdot 2$ we get an even order filter.
Weighing the kernel function
\begin{align}\label{eq:filter}
  h(n) = 
    \begin{cases}
      2\pi \cdot 2 \nicefrac{f_\text{nom}}{f_\text{s}} \,,                           & \text{for } n=\nicefrac{N}{2} \,,\\[1ex]
      \frac{\sin(2\pi \cdot 2 \nicefrac{f_\text{nom}}{f_\text{s}} \left(n-\nicefrac{L}{2} \right)}{n-\nicefrac{L}{2}}  \,,  & \text{otherwise} \\
    \end{cases} 
\end{align}
with a Hamming window results in a filter with its \SI{3}{\decibel} corner frequency close to $f_\text{nom}$. 
For \SI{25}{\kilo\hertz} sampling frequency and \SI{50}{\hertz} nominal frequency, this leads to a filter with $L = \text{100}$.

\begin{figure*}[t]
\begin{center}
\includegraphics[width=0.7\linewidth]{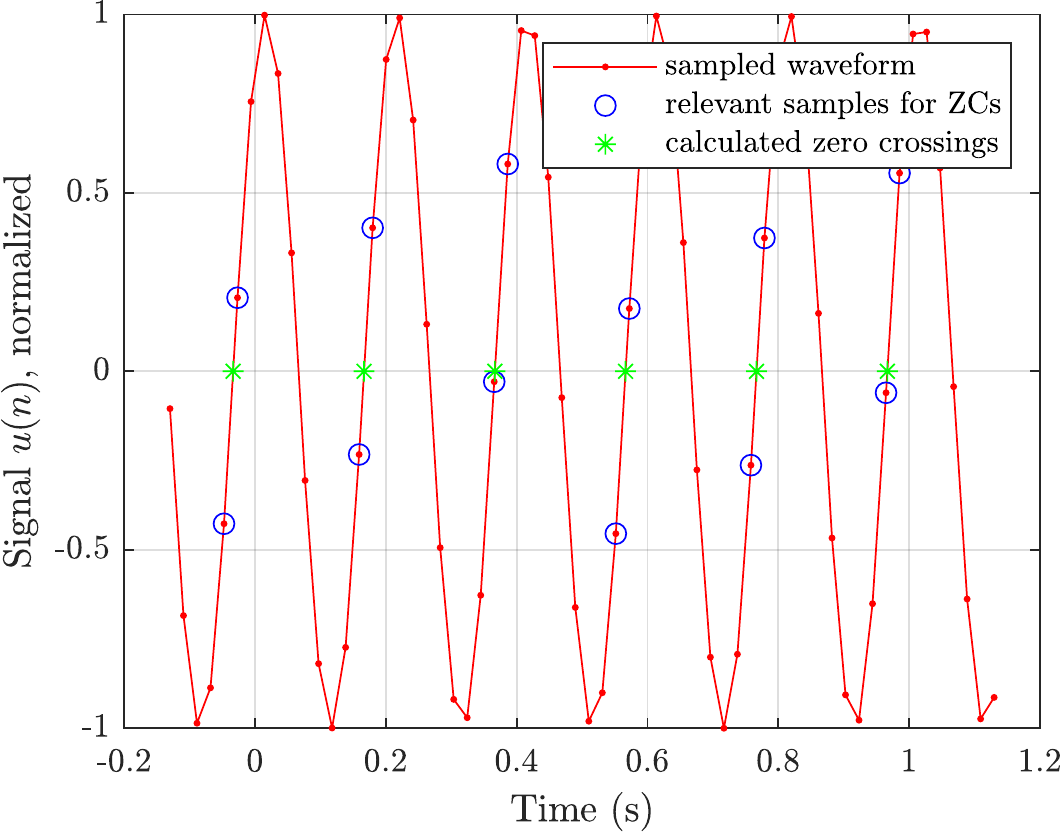}
\caption{We determine the power grid frequency from the captured samples as the reciprocal of the duration between rising zero-crossings (ZCs). 
Exact times of ZCs (asterisks) are calculated from the two adjacent samples (circles) to the ZC using linear interpolation. 
The last ZC before the beginning of a second (at $t=0$) is the first ZC included in the secondly average value. 
For this illustration we choose the signal frequency ($\SI{5}{\hertz}$) and sampling frequency ($\SI{50}{\hertz}$) much lower than in the actual acquisition system. } 
\label{fig:UTCsync}
\end{center}
\end{figure*}

Applying this filter to the samples $u(n)$ we obtain the filtered $\Tilde u(n)$ and calculate the current grid frequency by evaluating the time between two subsequent inclining ZCs of the waveform. 
Since the samples of the input signal will not coincide with actual point in time of the ZC, we linearly interpolate between two adjacent samples of a ZC to find the exact time of intersection with the zero line.
Figure~\ref{fig:UTCsync} shows in detail which samples are included in the process. 
This results in one frequency value $f(p)$ per period $p$ of the input signal.

Next, we have to choose an appropriate method of aggregation to provide one frequency reading per second.
Although trivial at first sight, aggregation has its challenge: 
The periods are usually not aligned with the interval given by a UTC second.
It is therefore important to consider the alignment of input signal and timestamp reference (i.e., the beginning of a second) to decide which periods $p$ belong to a aggregation interval $k$.
In our approach, the last inclining ZC before the start of a new second opens the interval.
The interval is closed again with the completion of the last full period within the second, i.e., the last inclining ZC before the next second.
This ZC is also the first for the following interval, and so on. 
The number of periods that are aggregated to form a secondly average therefore varies with the phase of the input signal and its frequency, see Table~\ref{tab:PeriodsPerSecond}.
%
\begin{table}[t]
    \caption{Number of aggregated measurements (periods) per second depending on the power grid frequency}
    \label{tab:PeriodsPerSecond}
    \centering
    \begin{tabular}{cc||cc}
    Frequency    &   Periods& Frequency    &   Periods\\
    \hline
    $\SI{49}{\hertz}\leq f < \SI{50}{\hertz}$     & 50, 51 & 
    $\SI{59}{\hertz}\leq f < \SI{60}{\hertz}$     & 60, 61\\
    $\SI{50}{\hertz}\leq f < \SI{51}{\hertz}$     & 49, 50 &
    $\SI{60}{\hertz}\leq f < \SI{61}{\hertz}$     & 59, 60\\
    \end{tabular}
\end{table}

\subsection{\revise{EDR hardware and setup}}
\label{sec:EDR_Hardware}
\begin{figure*}[t]
\begin{center}
\includegraphics[width=0.8\linewidth]{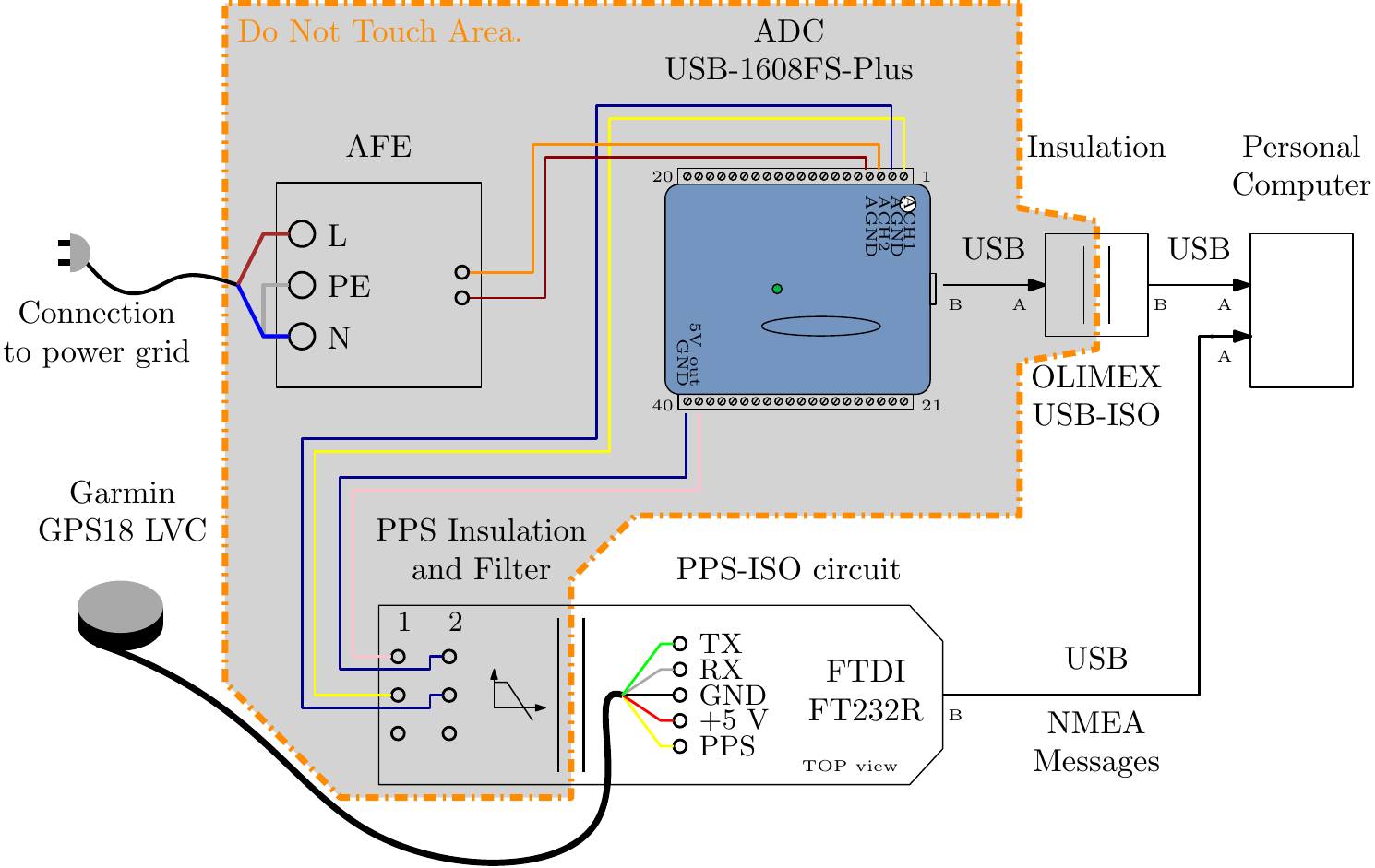}
\caption{
\revise{Hardware and interconnection details of the EDR. Parts that carry dangerous voltages (AFE, ADC, PPS-ISO, and USB-ISO) are to be placed inside an insulating enclosure. The shown wire colors are valid for the Garmin GPS18 LVC only, consult the user manual of your GPS receiver. Other interconnections are made with insulated $\leq \SI{0.25}{\milli\metre\squared}$ litz wire or ribbon cable.} }
\label{fig:EDR_HW_details}
\end{center}
\end{figure*}

\revise{
This Section provides additional technical details on the hardware and on setting up an EDR. Figure~\ref{fig:EDR_HW_details} illustrates how the logic function blocks of the EDR (shown earlier in Figure~\ref{fig:edrSystemOveriew}) are implemented in hardware components.
All design files (circuit schematics and printed circuit board layouts) are located in the data repository\cite{OSFdataBase} in the subfolder \url{Supplementary/HardwareDesignFiles/} and the source code of our acquisition software \texttt{EDR\_Scope\_N} is located in  \url{Supplementary/AcquisitionSoftware/}.
}

\paragraph{Disclaimer}
\revise{This is a technical description addressing persons with an appropriate electrical/electronic engineering background -- not a step-by-step tutorial. 
Experimenting with line-voltages can be dangerous and lethal if done incorrectly.
Do not attempt to rebuild the device unless you are familiar with construction of electronic circuits operating at line voltages, the associated dangers, and if required by law in your area, are certified to do experiments/build devices like this. 
Do not deploy self-built devices unless they have been individually tested and certified by the competent authorities in your area. 
We are not responsible for any damage that you cause or receive by following this description nor do we give any guaranties for the proper functioning of the things that you build.
}

\paragraph{Case}
\revise{
The highlighted area in Figure \ref{fig:EDR_HW_details} marks device parts that are directly connected to potentially dangerous voltage levels.
These parts are placed inside an insulating enclosure to prevent accidental or intentional touch. 
We use a Hammond 1599HGY case for this purpose.
}

\paragraph{AFE}
\revise{
The analog front end (AFE) scales down the line voltage to fit the ADC input range, using a voltage divider made from multiple metal film resistors. 
It provides basic safety measures by applying an overvoltage protection device (Dehn VC280 2).
The Neutral (N) input is tied together with the PE input, so that the circuit can be used without a Protective Earth (PE) connection. 
There is no insulation on this board, hence, the output might carry line potential measured against PE. 
}

\paragraph{PPS-ISO}
\revise{
Functions of PPS insulation, PPS filtering, and RS232-to-USB interface are combined into one custom made circuit that we call PPS-ISO.
The RS232-to-USB interface is built around an FTDI232RL chip. 
Drivers for this chip are integrated into modern operating systems, requiring neither special software nor any programming tools.
A MAX232 chip performs level conversion between the \SI{5}{\volt} logic and RS232 levels.
}

\paragraph{Connections}
\revise{
AFE and ADC as well as PPS-ISO and ADC need to be interconnected. 
The AFEs output connects to the second pair of input terminals (channel 2) of the ADC with insulated litz wire. 
Polarity is not important in this case since the polarity of the connection between power plug and socket is random in most of the settings.
Output pins 1 and 2 of PPS-ISO board are connected to pins 39 and 40 of the ADC module to supply the insulated side of the board with power from the PC.
PPS-ISO's output pin 3 carries the filtered PPS signal and connects to input channel 1 of the ADC while output pin 4 is the corresponding ground. 
The GPS receiver is wired to the screw terminal connector on the PPS-ISO board.
Pin assignments and wire colors (for Garmin GPS18 LVC) are according to Figure \ref{fig:EDR_HW_details} and can also be found in the design files. 
The TX and RX connections need to be crossed over, as depicted.
Hence, the TX wire from the GPS receiver (gray/white) must be connected to the RX terminal of the PPS-ISO and the the RX wire of the receiver (green) needs to be connected to the TX terminal. }

\revise{
Using standard USB-A to USB-B cables, the ADC connects to the OLIMEX USB-ISO and the USB-ISO to the PC. 
Likewise, the PPS-ISO is connected to the PC. 
}

\paragraph{Software preparation}
\revise{
Communication with the ADC module (USB1608FS-Plus) requires the installation of a proprietary driver suite from Measurement Computing GmbH\footnote{Usually this software is shipped with the ADC module. Otherwise consult Measurement Computing GmbH's website \url{https://www.mccdaq.de}}. 
Once the ADC is connected to the PC, the \texttt{InstaCal} executable from Measurement Computing GmbH needs to be run once to identify and enlist the ADC module. 
Starting the \texttt{EDR\_Scope\_N} executable initiates the data acquisition process and file generation.
Using a configuration file (\texttt{EDR-Scope.ini}), key parameters like sampling rate and data tags are set.
The executable can be compiled using the provided Microsoft Visual Studio project, which contains all required source code.
}

\subsection{Code availability}
\revise{
The source code for the acquisition executable \texttt{EDR\_Scope\_N}, the schematic diagrams and hardware design files, 
\textsc{Matlab} functions for calculating the frequency time series from synchronized samples, and the scripts for generating the graphics and analyses in this paper are freely available along with the data\cite{OSFdataBase} in the subfolder \url{Supplementary/}. 
We use \textsc{Matlab} Version 2020a.
Hardware design files are compatible with Autodesk EAGLE6.3 and upwards.}

\section{Data records}
\label{sec:DataRecords}
The file-based data records are available online under \url{https://osf.io/by5hu/} \cite{OSFdataBase}.
We provide two subfolders: \emph{FrequencyData} containing the frequency data in one-second resolution; 
and \emph{WaveformData} containing excerpts of the sampled signal.

We use a two-letter code (like DE for Germany) in accordance with the ISO 3166 to designate the regions in which the measurements were taken.
Table~\ref{tab:abbreviations} maps the abbreviations to the campaigns, the regions, and the files.
In some regions we conducted multiple campaigns in different locations.
These are differentiated by adding a sub region code specific to the location.
For example, data recorded in Texas (United States of America) are identified by the key US\_TX.
As we intend to continue data collection, the filenames also carry an index as campaign identifier.
An example filename construction looks like this:
\begin{figure*}[ht]
\centering
\includegraphics[width=0.4\linewidth]{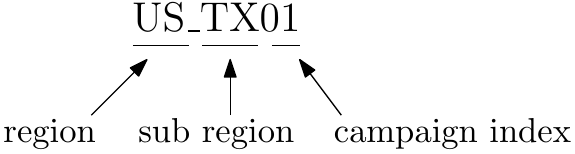}
\end{figure*}

\subsection{Frequency data}
For the isolated measurements (e.g. on islands) we provide one file per campaign, including timestamp and frequency trajectory.
Files of measurements within the Continental European synchronous area (EC) also contain the frequency measured in Karlsruhe together with a common timestamp.
The data of the synchronized campaign including the measurements from Lisbon (Portugal), Oldenburg and Karlsruhe (Germany), as well as Istanbul (Turkey) are located in a single file (SYNC01.csv) using a common timestamp.

\begin{table}[ht]
\caption{Data format of CSV files containing the frequency time series. Delimiter: Semicolon. Total number of columns depends on data set.}
\label{tab:CSVdesc}
\begin{tabular}{cp{1.5cm}p{4cm}p{4cm}}
\hline
\multicolumn{1}{c}{Col.} & \multicolumn{1}{l}{Head} & \multicolumn{1}{l}{Description} & \multicolumn{1}{l}{Format} \\ \hline
1 & Time & Date and time, UTC & yyyy-MM-dd HH:mm:ss \\
2 & f50\_'loc' & $f - f_\text{nom}$ at 'loc' (\si{\milli\hertz}) & decimal: -123.456 \\
3 & QI\_'loc' & Quality indicator for 'loc' series & integer ($0\ldots3$) \\ \hline
4 & f50\_'loc' & $f - f_\text{nom}$ at 'loc' (\si{\milli\hertz}) & decimal: -123.456 \\
5 & QI\_'loc' & Quality indicator for 'loc' series & integer ($0\ldots3$) \\ \hline
\end{tabular}
\end{table}

Frequency data are provided as CSV of the format shown in Table~\ref{tab:CSVdesc}.
We use a semicolon delimiter.
The frequency columns are marked with `f50' or `f60' according to the nominal frequency in the measurement location, i.e., \SI{50}{\hertz} or \SI{60}{\hertz}, respectively.
We always report the difference of the measured frequency $f$ and nominal frequency $f_\text{nom}$ in \si{\milli\hertz}.
So f50 is
\begin{equation}
\text{f50} = (f - \SI{50}{\hertz})\times10^{-3}    
\end{equation}
To ease (partial) data import, each line carries a full timestamp.
The timestamps are guaranteed to be regular, so they are equidistant with one second spacing and unique. 
Therefore, they may be omitted to speed up the import.
Besides the timestamp, all data are numbers using a period as decimal limiter. 

Data points that were corrupted during the acquisition, e.g. by the loss of GPS reception, are linearly interpolated.
The user can distinguish these points by evaluating the column of the quality indicator (QI) described in Table~\ref{tab:QIdescription}.

\begin{table}[ht]
\centering
\caption{Quality indicator (QI) for frequency data}
\label{tab:QIdescription}
\begin{tabular}{c|l}
\hline
QI & Description \\ \hline
0 & Data OK (as specified) \\
1 & Invalid data (only in synchronous campaign) \\
2 & Interpolated data \\
\end{tabular}
\end{table}

In the Continental European synchronized campaign (Karlsruhe (Germany), Oldenburg (Germany), Turkey, Portugal) not all frequency recordings begin and end at the same time.
The file begins with the start of the first recording and stops with end of the last recording.
Consequently, there are times for which some of the data are not available.
At these times, entries in the affected frequency columns are filled with zeros and are marked as ``invalid data'' in the corresponding quality indicator column. 

\renewcommand{\arraystretch}{1.2}{
\begin{table*}
\caption{Keys of measurement locations and their connection to synchronous areas. 
For each country, we adopt the ISO 3166 code and further specify the region or city when multiple measurements were recorded in the same country. 
Devices are connected at power sockets, except for DE\_KA. PCC stands for point of common coupling.}
\label{tab:abbreviations}
\begin{centering}
\begin{tabular}{p{1.2cm}p{1.5cm}p{4cm}p{2.1cm}p{1.3cm}}
\hline 
Key &  File names & Measurement location & Synchronous area & Environ- ment\tabularnewline
\hline 
\multicolumn{5}{c}{\textbf{Islands}}\tabularnewline
\hline 
IS & IS01 & Reykjavík, Iceland & Iceland & Hotel \tabularnewline
FO & FO01 & Vestmanna, Faroe Islands & Faroe Islands & Hotel\tabularnewline
ES\_GC & ES\_GC01 ES\_GC02 &  Las Palmas de Gran Canaria, Canary Islands, Spain & Gran Canaria & Hotel\tabularnewline
ES\_PM & ES\_PM01 & Palma de Mallorca, Balearic Islands, Spain & Mallorca & Office \tabularnewline
\hline 
\multicolumn{5}{c}{\textbf{Continental}}\tabularnewline
\hline 
DE\_KA & SYNC01 & Karlsruhe, Germany & Continental Europe (CE) & Office, PCC\tabularnewline
DE\_OL & SYNC01 & Oldenburg, Germany & CE & Home\tabularnewline
TR & SYNC01 & Istanbul, Turkey & CE & Office\tabularnewline
PT & SYNC01 & Lisbon, Portugal & CE & Home\tabularnewline
FR & FR01 & Lauris, France & CE & Home\tabularnewline
HR & HR01 & Split, Croatia & CE & Hotel\tabularnewline
PL & PL01 & Krakau, Poland & CE & Hotel\tabularnewline
IT & IT01 & Erice, Sicily, Italy & CE & Hotel\tabularnewline
GB & GB01 \newline GB02 & London, United Kingdom & Great Britain & Home\tabularnewline
EE & EE01 & Tallinn, Estonia & Baltic & Office \tabularnewline
SE & SE01 & Stockholm, Sweden & Nordic & Hotel\tabularnewline
\hline 
\multicolumn{5}{c}{\textbf{Others}}\tabularnewline
\hline 
US\_UT & US\_UT01 & Salt Lake City, Utah, US & Western Interconnection & Hotel \tabularnewline
US\_TX & US\_TX01 US\_TX02& College Station, Texas, US & Texas Interconnection & Hotel \tabularnewline
ZA & ZA01 & Cape Town, South Africa & South Africa & Hotel\tabularnewline
RU & RU01 & St. Petersburg, Russia & Russia & Hotel\tabularnewline
\end{tabular}
\end{centering}
\end{table*}
}

\subsection{Waveform samples}
Despite the focus of the data base on frequency data, we also provide excerpts of the sampled waveform of the voltage at the connection point of the EDR. Each excerpt contains an one hour recording at \SI{25}{\kilo\hertz} sampling frequency from all synchronous areas. 
Data are provided as a CSV file with one voltage entry per time step, which is $\nicefrac{1}{f_\text{s}} = \SI{40}{\micro\second}$.
To reduce overhead, the files do not contain timing information.
Instead, the file name carries the timestamp of the first sample in the file, from which the user may reconstruct the time vector.

The waveform samples are influenced by local phenomena in the vicinity of the recorder (switching operation of appliances etc.) over which we had no control and have no additional information.
The value of the waveform samples might therefore be limited and care must be taken when interpreting and working with them.
Due to the large data amount we choose to only provide excerpts.
However, extended waveform data are available upon request.

It is important to note that the polarity and phase (either L1, L2, or L3 in three-phase power systems) of the signal should be considered unknown.
If the absolute phase angle matters, the user might have to correct the phase using a local reference.
In location DE\_KA we assume be connected to phase one (L1) in correct polarity, but no guarantees can be made.

\section{Technical validation}
\label{sec:TechnicalVal}
In the following sections we present methods and results for the validation of this data set.
We begin by comparing the recorded data from multiple instances of our instrument with each other and with external sources to quantify the error in the typical recording conditions. 
Subsequently, errors and limitations of our frequency calculation algorithm are discussed in a broader form.
This discussion is motivated by the fact that there is no general concept to \revise{estimate the instantaneous frequency of a distorted (non-sinusoidal) signal without error.}
We close with remarks on the data quality in terms of total data amount and share of missing data. 

\subsection{Comparative frequency measurements}
\label{sec:ComparativeExp}
To evaluate the performance of our frequency measurements in actual measurement conditions, we prepare a comparative setup.
Two EDR instruments are placed next to each other with their GPS receivers located in close proximity. 
They are connected to the same power socket but plugged in opposite polarities.
Using this setup we record the grid frequency simultaneously for approximately $10$ hours with both devices.
We get two frequency time series, $f_1$ and $f_2$.
The recording interval is from `2019-06-27 20:10' to `2019-06-28 06:50' and its data are available along with the data set. 
Using the timestamp we compare both recordings. 
Rows with missing data are discarded.

\begin{figure*}[hp]
\includegraphics[width=0.95\linewidth]{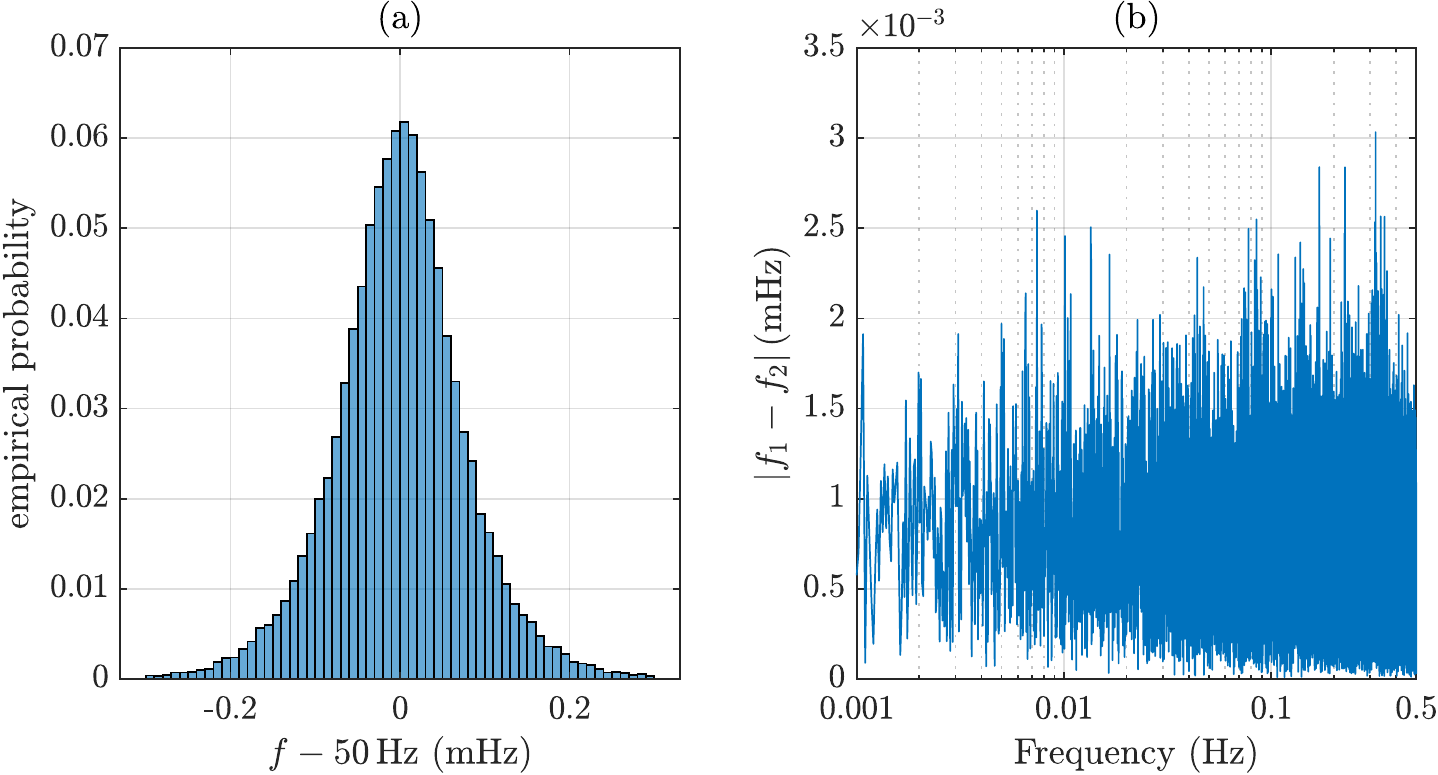}
\caption{Comparing the frequency records of two EDR instruments from the same location: a) Histogram of the frequency differences. b) Single-sided amplitude spectrum of the frequency differences.} 
\label{fig:ErrorHistAndSpec}
\end{figure*}

Figure~\ref{fig:ErrorHistAndSpec} (a) depicts the empirical error distribution from this test.
We find a mean error of \SI{-34}{\nano\hertz} and an RMSE of \SI{78}{\micro\hertz}.
The maximum deviation in the experiment is \SI{443}{\micro\hertz}, the median \SI{45}{\micro\hertz}, and the $99^{\mathrm{th}}$ percentile \SI{236}{\micro\hertz}.
Analyzing the deviations $f_1-f_2$ in frequency domain shows several distinct peaks and an increase in amplitude towards higher frequencies, see Figure~\ref{fig:ErrorHistAndSpec} (b).
None of the peaks relates to any known process or process cycle within the acquisition or signal processing. 
Since the amplitude of the individual spectral components is small compared to the RMSE, the error is not dominated by a single periodic mechanism.

\subsection{Comparison with external data source}
In the following we compare data from an external source from the same synchronous region with our recordings.
This experiment intends to show any temporal alignment missmatches and to give a general idea of how expected differences to other sources might look like. 
Deriving measurement accuracy from this experiment is not advisable since a) the measurement locations differ and b) we have no knowledge about how and to what accuracy the external system performs.
Physical effects in the power distribution add indistinguishably to the measurement errors.

For the Continental European synchronous area the french TSO Réseau de Transport d'Electricité (RTE) provides frequency times series in $10$-second resolution \cite{RTE-UCTE2016}. 
We compare the RTE data to our own measurements by aggregating our data also to a $10$ second resolution.
We use the same data as for the EDR-to-EDR comparison in Section~\ref{sec:ComparativeExp}. 
Since the timestamp of the RTE data is aligned in the center of the aggregation window, we have to adjust our time series to match this scheme.

\begin{figure*}[h!]
\includegraphics[width=0.95\linewidth]{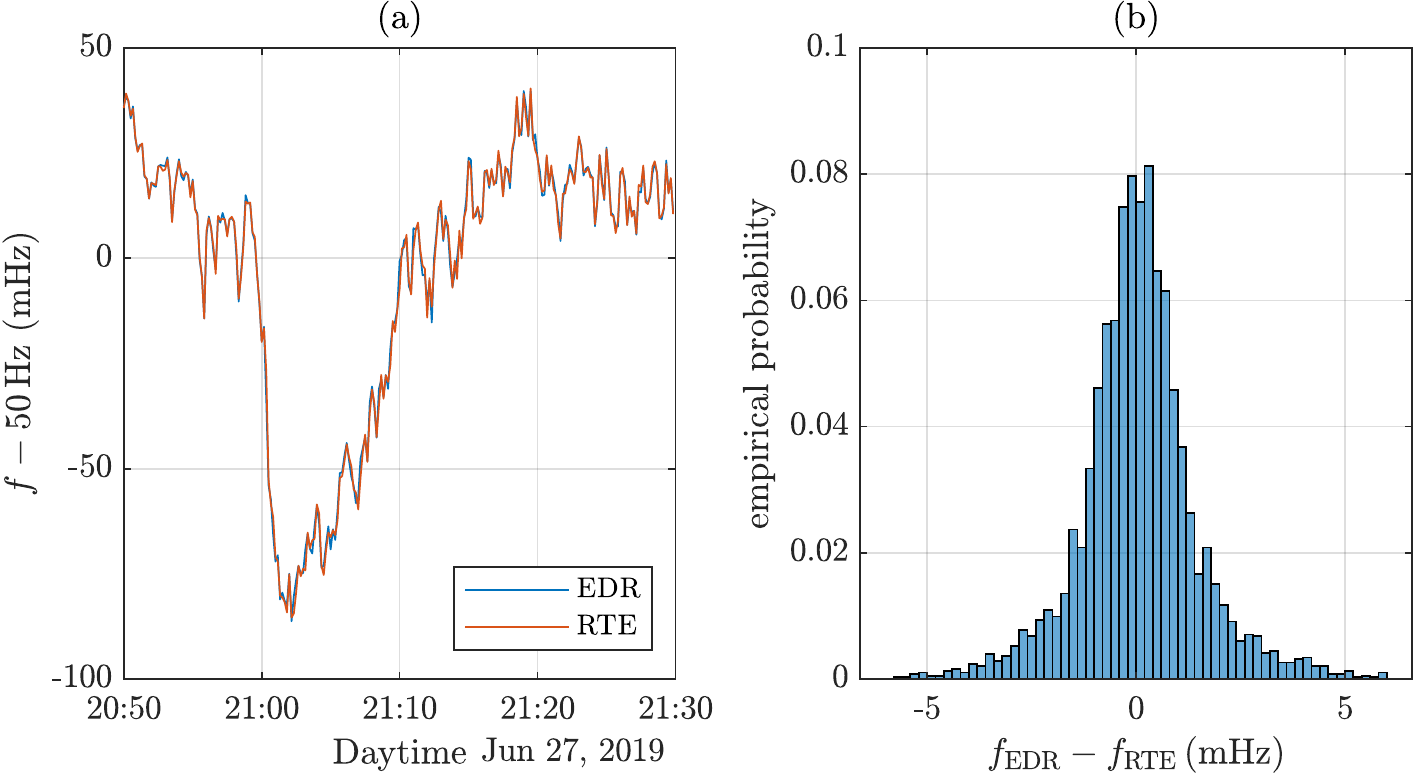}
\caption{Comparison of EDR and RTE measurements for a $10$ hour recording: a) Visual comparison of an excerpt, showing a close match. b) Histogram of the differences showing the error distribution.} 
\label{fig:EDRvsRTE}
\end{figure*}

Figure~\ref{fig:EDRvsRTE} compares the two frequency time series.
Figure~\ref{fig:EDRvsRTE} a) depicts an excerpt of both time series, which shows that the trajectories match well on visual inspection.
Calculating the sample by sample difference leads to an RMSE of \SI{1.57}{\milli\hertz}. 
The histogram of the deviations is depicted in Figure~\ref{fig:EDRvsRTE} b).
The error is roughly an order of magnitude larger than what we have observed directly comparing two of our instruments.
This may be caused by the reasons indicated above.

\subsection{Discussion on the frequency estimation technique}
\revise{The most basic sinusoidal signal model for the power grid waveform is given by: 
\begin{equation}
    u(t) = A_\text{nom}\sin\left(2\pi f(t)\,t + \varphi(t)\right),
\end{equation}
with the power grid frequency $f(t)$, a phase angle $\varphi(t)$ and the amplitude $A_\text{nom}$.
During grid operation $f(t)$ varies slightly around the nominal power grid frequency $f_\text{nom}$ in a band defined by the grid control strategy. 

The ZC algorithm estimates the frequency $f(t)$ by finding the zero-crossing points and the time span between them.  
$f(t)$ is then calculated as the inverse of this time span.
This procedure is a sampling process since the signal and therefore the frequency, is \emph{probed} at specific points in time only.}
There are four major aspects that influence the quality of the frequency estimation using the EDR and the presented ZC algorithm. These are:
\begin{enumerate}
    \item Disturbances on the power signal
    \item Phase sensitivity of frequency calculation
    \item Temporal alignment between instruments 
    \item Stability of the clock reference
\end{enumerate}

\paragraph{Disturbances on the power signal}
Disturbances on the power signal cause the input waveform to differ from a sine, which leads to errors in determining the proper moment when the signal crosses zero.
Static harmonics of any order distort the waveform, but they do not alter the length of periods.
As most of the harmonic content is filtered out during signal processing, it does not influence the accuracy of the estimation. 

On the contrary, any random signal that passes the filtering will cause frequency estimation errors, also on the aggregated level.
To investigate this influence, we synthetically generate sinusoidal signals with superimposed white noise of different power levels.
We use a signal length of one hour and pass the samples through the processing chain, shown in Figure~\ref{fig:dataflow}.

\begin{figure}[ht]
    \centering
    \includegraphics[width=0.7\linewidth]{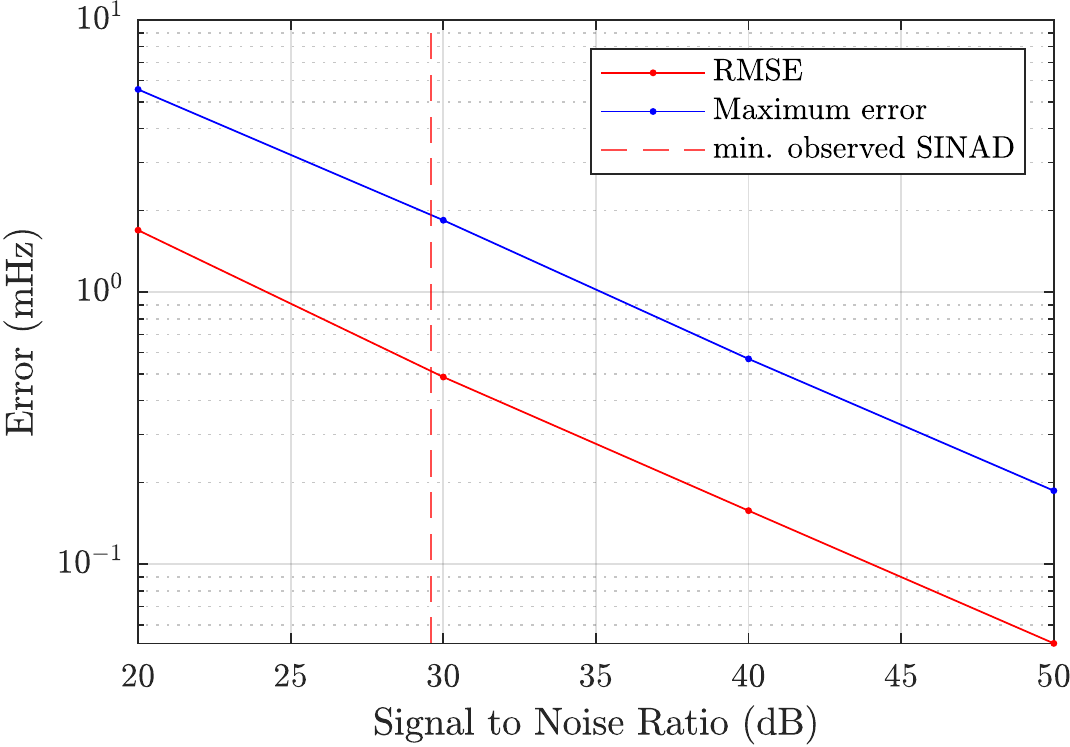}
    \caption{Deviation of the ZC frequency estimate due to additive white noise in signal input. We use $f_\text{nom}=\SI{50}{\hertz}$ as input frequency and one-hour of synthetic data. Vertical line: Lowest Signal to Noise and Distortion ratio (SINAD) observed in the measurement campaigns (SE01) = \SI{29.6}{\decibel}.}
    \label{fig:NoiseSensitivity}
\end{figure}

Figure \ref{fig:NoiseSensitivity} depicts the observed maximum error and RMSE of the frequency estimate over the Signal to Noise Ratio (SNR).  
The vertical line indicates the lower limit of the Signal to Noise and Distortion Ratio (SINAD) over all measurement campaigns. 
In this setting, we count all distortions as noise although the harmonics dominate in the measurement campaigns.
The rationale is to apply the reproducible albeit harder test case.

The test results in an maximum error of \SI{2}{\milli\hertz} and an RMSE of \SI{0.5}{\milli\hertz} in the given measurement scenarios.
Thus, signal distortion is one of the major contributors to the total error in relation to the following error sources. 

\paragraph{Phase sensitivity of frequency calculation}
The ZC algorithm relies on detecting inclining ZCs.
Therefore, the absolute phase (phase with respect to UTC) of the signal determines which of the period-wise frequency measurements $f(p)$ are aggregated to form a one-second-reading $f(k)$.
This behavior represents an error in the sense that two instruments can report different frequencies depending on the phase of the signal.
The difference is a direct consequence of the non-continuous definition of the frequency used in the ZC algorithm.
This effect occurs, for instance, if one instrument is plugged into a power socket reversely compared to another.
The series of periods that is included into $f(k)$ by this instrument begins half a period later or earlier ($\pm\SI{10}{\milli\second}$ in a \SI{50}{\hertz} system) compared to the other one.
This misalignment results in a frequency error when comparing the two aggregated frequency measurements from those two instruments.

The difference between the \emph{true} one-second average of the grid frequency 
and the estimation $f(k)$ is hard to quantify, as it depends on the momentary phase of the input signal.
We approach this issue by running a simulation for which we generate artificial samples that we pass through our signal processing chain (Figure~\ref{fig:dataflow}). Via a frequency modulator, we can generate a realistic but synthetic signal
for which we know the frequency exactly.
Further, shaping the spectrum of the modulation signal allows to investigate the influence of spectral components of the frequency changes on the precision of the estimation.

We begin by defining the momentary or instantaneous frequency $f_\text{i}(t)$ of a sinusoidal signal
\begin{equation}
    f_\text{i}(t) = f_\text{nom} + \frac{1}{2\pi}\,\frac{\text{d}}{\text{d}t} \varphi(t)\,.
\end{equation}
A frequency modulation is characterized by
\begin{equation}
    f_\text{i}(t) - f_\text{nom} = D_\text{f} \, m(t)\,,
\end{equation}
where $D_\text{f}$ is the modulation depth and $m(t)$ some modulation signal.
To apply frequency modulation, i.e., to change $f_\text{i}(t)$ in dependence of $m(t)$, the phase $\varphi(t)$ has to be
\begin{equation}
\label{eq:fmod}
    \varphi(t) = 2\pi D_\text{f} \int_{-\infty}^{t} m(\tau) \text{d}\tau\,.
\end{equation}
 As we are operating in discrete time domain, we generate samples of a frequency modulated band pass signal by using the trapezoidal approximation for the integral in \eqref{eq:fmod}.
The frequency modulated test signal $u_\text{test}(n)$ is therefore
\begin{equation}
    u_\text{test}(n) = A_\text{nom}\sin\left(\frac{n}{f_\text{s}} 2\pi  f_\text{nom} + 
    2\pi D_\text{f} \frac{1}{2} \sum_{\theta=1}^{n}\frac{m(\theta-1)+m(\theta)}{f_\text{s}}
    \right).
\end{equation}
To judge the quality of the one-second average estimate $f(k)$, we define the true one-second average of the frequency of the signal $u_\text{test}(n)$ as
\begin{equation}
    \bar{f}_\text{test}(k) = \frac{1}{N}\sum_{n=kN+1}^{(k+1)N} f_\text{nom} + D_\text{f}\,m(n) \qquad \text{for } k = 0,1,2,\ldots\,,
    \label{eq:trueAverage}
\end{equation}
which is the secondly average \emph{regardless} of the current phase of the signal.

\begin{figure}[h]
    \centering
    \includegraphics[width=0.99\linewidth]{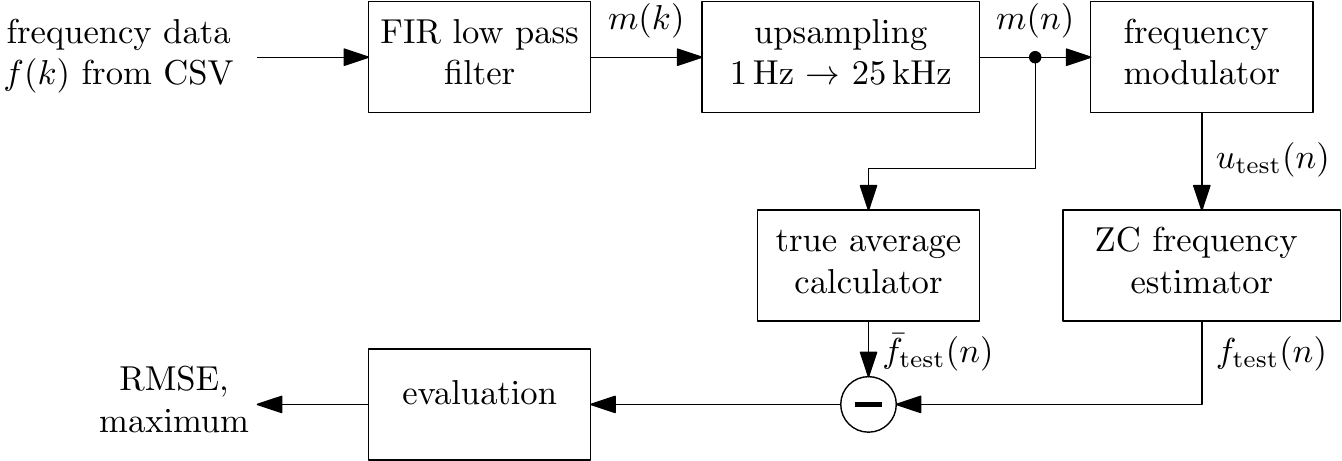}
    \caption{Logical representation of the experimental setup to estimate the error of the ZC algorithm due to phase sensitivity.}
    \label{fig:PhaseSensExpSetup}
\end{figure}

\begin{figure}[hp]
    \centering
    \includegraphics[width=0.99\linewidth]{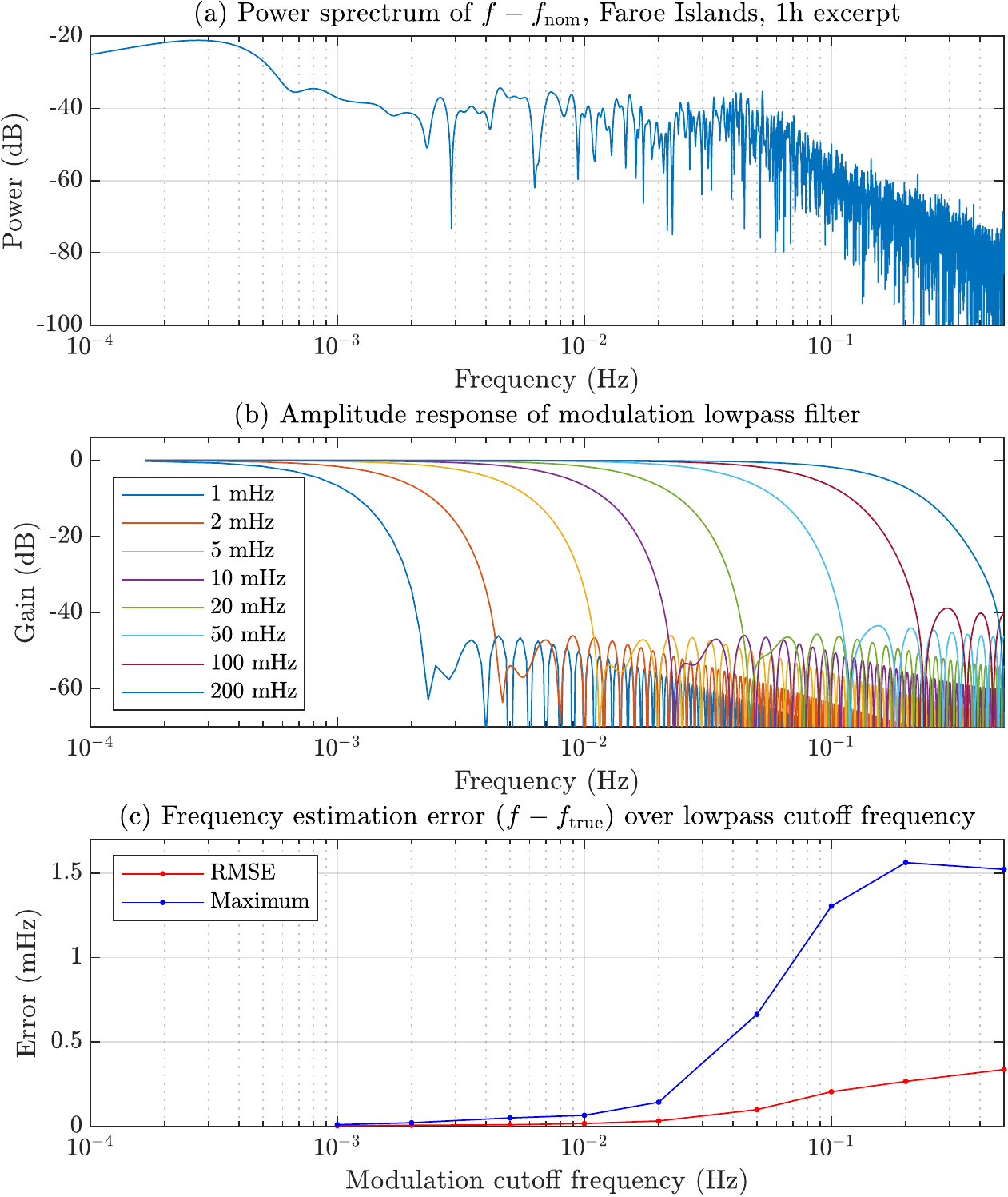}
    \caption{(a) Power spectrum of the frequency deviations, as reported in the CSV files. Here the deviations serve as modulation signal. Spectrum generated using a $2^{16}$-point FFT on an one-hour excerpt from the Faroe Island data. (b) Amplitude frequency responses of the FIR low-pass filters used on the modulation signal. Note the constant roll-off steepness for the different cutoff frequencies. (c) Deviation between the true one-second average and the estimated frequency (using the zero-crossing algorithm) in dependence of the low-pass cutoff frequency when applied to a sinusoidal test signal that is frequency modulated with the output of the low-pass filter.}
    \label{fig:PhaseSensitivity}
\end{figure}

The process (depicted in Figure \ref{fig:PhaseSensExpSetup}) comprises to take a frequency trajectory (as provided in the CSV files), optionally pass it through a low-pass filter, and use it as $m(n)$ to generate an artificial signal with the same frequency trajectory.
Since $m(n)$ is quasi-continuous we can obtain the true one-second frequency average according to \eqref{eq:trueAverage}.
These averages are subsequently compared with the results from the ZC algorithm.
We calculate the RMSE and maximum deviations over the length of the artificial signal (one hour in our case).
Figure \ref{fig:PhaseSensitivity} presents the results of this study. 
Figure~\ref{fig:PhaseSensitivity} (a) depicts the power spectrum of the frequency.
We choose the data set from the Faroe Islands for its high modulation signal power.
Using a low-pass filter we suppress some of the higher frequency content from $m(n)$, while preserving the total signal power. 
Figure~\ref{fig:PhaseSensitivity} (b) shows the amplitude filter responses.
We choose the filters to be FIR with decreasing length, so that the roll-off always has the same steepness.
This prevents cross-over in the transition region.
Finally, Figure~\ref{fig:PhaseSensitivity} (c) displays the resulting errors in dependence of the chosen cutoff frequency for the low-pass filter. 
The rightmost point (\SI{0.5}{\hertz}) relates to no filtering and is therefore equivalent to the setting of the measurement campaign. 
Thereby, the maximum error is \SI{1.52}{\milli\hertz} and the RMSE is \SI{0.34}{\milli\hertz}.
Increasing the cutoff frequency results in larger error which corresponds to the susceptibility of the ZC algorithm to sudden frequency changes.

The error figures for signal distortion and for phase sensitivity are in the same range ($\text{RMSE} \approx \SI{0.5}{\milli\hertz}$) and are therefore considered to contribute equally to the total deviations in the data set.

\paragraph{Temporal alignment between instruments}
When comparing aggregated data, the alignment of the underlying sampled time series is critical.
Sampling the PPS within the acquisition system results in a static delay (phase offset) due to tolerances of the used electrical components.
Therefore, there is a phase shift between the internal clock and UTC. 
This error does not manifest itself when investigating the frequency data from a single instrument, but shows up when comparing data from two sources. 
The impact of this delay on the frequency time series is dependent on the measured signal.
In steady state condition, the effect vanishes and it has increasing influence when the frequency changes rapidly. 

We determined that the device specific and static timing offset is $\pm \SI{5}{\micro\second}$ (3$\sigma$)\cite{EDR_Rework2019}.
Two instruments might therefore be misaligned by $\SI{10}{\micro\second}$ or one quarter of a sample at $\SI{25}{\kilo\hertz}$ sampling frequency.
As this misalignment is very small compared to the aggregation interval of the frequency time series (one second), it is only marginally contributing to the total error.

\paragraph{Stability of the reference clock}
The stability of the reference clock is key to keep offsets in the frequency measurements to a minimum.
As the frequency is calculated from the time between zero-crossings, the corresponding frequency error $\Delta f$ to a given uncertainty in the time $\Delta t$ is
\begin{equation}
    \Delta f = \left|-(t=\nicefrac{1}{f_\text{nom}})^{-2}\right|\Delta t.
\end{equation}

The uncertainty in the clock relates to the interval of one period of the input waveform. 
We use the highly stable PPS signal \cite{PPS_Qual} with $\pm \SI{1}{\micro\second}$ accuracy to determine the duration of a second in our instrument.
Experimentally, we find that the error in steady state conditions over a 10 minute interval is \SI{26.7}{\micro\hertz} at \SI{50}{\hertz} within a 95.4\% confidence interval \cite{EDRprepForComp}.
This error is therefore irrelevant compared to the previously discussed error sources.

\subsection{Data quality}
Despite our greatest efforts, the time series in this data set contain multiple missing entries.
Some campaigns are fragmented because we conducted measurements during business or private travels.
Other interruptions were caused by a loss of GPS reception.
For the majority of short interruptions (<~\SI{10}{\second}) we believe that they originate from acquisition process interruptions caused by the operation system on our device. 
The share of fully trustworthy data is \SI{97.6}{\percent} overall.
This number is probably lower than it could be: 
We decided to be very strict by considering each one-second aggregate as faulty if at least one period measurement was missing.
In Table~\ref{tbl:FilesAndProperties} we provide basic information on the campaigns as well as numbers for the share of missing data for each file.  

\begin{landscape}
\renewcommand{\arraystretch}{1.1}{
\begin{table}[ht]
\centering
\caption{Measurement campaign files and properties}
\label{tbl:FilesAndProperties}
\begin{tabular}{lllSSSl}
\hline
File    &   Begin   & End   &   {Duration~(d)}    &   {Missing~(\%)} &  {Cont. Days} &   Cont. Reg. Begin \\ \hline
\multicolumn{7}{c}{\textbf{Islands}}\tabularnewline
IS01    & 2017-10-14 17:23:45  & 2017-10-20 07:54:06  & 5.6  & 24.0  & 0.87  & 2017-10-18 16:27:12\\
FO01    & 2019-11-03 21:50:10  & 2019-11-10 09:15:06  & 6.5   & 33.3  & 0.03  & 2019-11-04 05:01:20\\
ES\_GC01& 2018-02-04 10:36:57  & 2018-02-10 21:28:57  & 6.5 & 0  & 6.45  & 2018-02-04 10:36:57\\
ES\_GC02&  2018-11-25 00:00:00 & 2018-11-26 12:29:44  & 1.5  & 42.2  & 0.28  & 2018-11-25 00:00:00\\
ES\_PM01   &2019-09-29 00:00:00   & 2019-12-31 23:59:59  & 94.0  & 0.02   & 4.7  & 2019-11-27 10:17:19\\
GB01    &  2019-03-04 10:54:37  & 2019-03-07 23:59:59  & 3.5  &   1.96 &   1.0 & 2019-03-06 00:00:00\\
GB02    & 2019-11-10 22:36:18  & 2019-12-31 23:59:58  &  51.1 & 0.2   &  3.8 & 2019-12-25 07:27:20\\
\multicolumn{7}{c}{\textbf{Continental synchronous areas in Europe}}\tabularnewline
SYNC01  &   &   &   &   &   & \\
 - DE\_KA & 2019-07-09 00:00:00  &  2019-08-18 23:59:59 &  41.0 & 0.0  &  41.0 &  2019-07-09 00:00:00\\
 - DE\_OL & 2019-07-10 11:09:02  & 2019-08-07 16:14:00  & 41.0  & 0.05  &  14.7 & 2019-07-11 21:30:37\\
 - PT    & 2019-07-09 19:34:40  &  2019-08-18 09:52:21 &  41.0  & 4.4  & 5.4  & 2019-08-01 23:30:46\\
 - TR    & 2019-07-09 21:01:29  & 2019-08-16 14:12:59  &  41.0 & 0.01  & 3.7  & 2019-07-31 23:59:55\\
FR01    & 2019-04-16 00:00:00   & 2019-04-27 06:56:17   & 12.0    &  1.74 & 0.24  & 2019-04-19 05:03:47 \\
HR01    & 2019-04-09 14:31:34   & 2019-04-12 09:47:02   & 4.0     &  0.0  & 1.25  & 2019-04-10 19:12:54 \\
IT01    & 2019-07-02 20:32:32   & 2019-07-06 16:46:21   & 5.0     &   5.4 & 0.7   & 2019-07-05 09:03:29 \\    
PL01    & 2019-04-04 13:05:53  & 2019-04-07 10:42:55  & 4.0  & 0.75  & 0.66  & 2019-04-05 03:22:29\\
EE01    & 2019-03-25 11:09:11  & 2019-04-17 07:32:47  & 22.9  & 0.0  & 2.0  & 2019-03-29 00:00:00\\
SE01    & 2019-05-06 14:16:09  & 2019-05-13 06:54:20  & 6.7  & 2.15  & 0.82  & 2019-05-09 20:34:30\\
PT01    & 2018-02-14 14:33:09   & 2018-02-21 09:10:00 &   6.8 &   0.17& 0.23  & 2018-02-21 01:43:28\\   
\multicolumn{7}{c}{\textbf{Other}}\tabularnewline
US\_UT01    & 2019-05-19 02:54:06  & 2019-05-25 11:58:16  & 6.4  & 0.4 & 2.29  & 2019-05-23 04:58:00\\
US\_TX01     & 2019-05-15 04:35:36  & 2019-05-16 15:03:56  & 1.4  &  0.18 & 0.6 & 2019-05-15 04:35:36\\
US\_TX02     &  2019-05-20 07:21:15 &  2019-05-23 23:59:58 & 3.7  & 4.04  & 0.61  & 2019-05-20 07:28:44\\
ZA01    & 2017-11-19 11:33:08  & 2017-11-28 23:59:59  & 9.5  & 33.6 & 0.89  & 2017-11-22 03:43:16\\
RU01    & 2019-04-30 23:00:17  &   2019-05-12 07:37:57 & 13.0  & 0.01  & 1.3  & 2019-05-04 10:23:32\\
\hline
Total:  &   &  & 428.1  & 2.4  &   & \\
\end{tabular}
\end{table}
}
\end{landscape}

\section{Usage notes}
In the following, we provide some hints for working with the data.

\subsection{Import of frequency data}
As the frequency time series are provided as CSV, they can be easily directly imported with a variety of software tools. 
Below we provide code snippets for commonly used programming languages to import the data. 
After import, one should check that the data were read correctly (with the available precision) by plotting or printing the first couple of entries.
The user must take care of the correct interpretation of the timestamp, especially if the data are combined from different sources. 
This data set solely uses UTC timestamps while data from external sources (e.g. TSOs) commonly use local time. 
\begin{lstlisting}[title={Matlab:}]
%since Matlab Version 2020a, direct import to timetable
Data = readtimetable('GB01.csv');

%before e.g. like this
df = importdata("GB01.csv")
Time = datetime(df.textdata(2:end,1));
header = df.textdata(1,:);
data = df.data;
\end{lstlisting}

\begin{lstlisting}[title={Python 3 (using Pandas):}]
import pandas as pd 
Data = pd.read_csv("GB01.csv") 
\end{lstlisting}
\begin{lstlisting}[title={R:}]
library(zoo)
DF <- read.zoo("GB01.csv",
               tz = "UTC",
               FUN = paste,
               index = "Time",
               FUN2 = as.POSIXct,
               sep = ";",
               header = TRUE,)
\end{lstlisting}

\subsection{Restore scale and frequency offset}
The data are provided in \si{\milli\hertz} without the nominal frequency offset. If the application requires the frequency to be in \si{\hertz} and to include the nominal frequency offset, one needs to restore it by calculating
\begin{equation}
    f=\text{f50\_LOC}\times 10^{-3} + f_\text{nom}.
\end{equation}
Thereby, $\text{f50\_LOC}$ are the entries in frequency column in a data set with \SI{50}{\hertz} nominal frequency. 
One can obtain the nominal frequency of the selected region from the head of the frequency column in CSV file, which reads \emph{f50} or \emph{f60} depending on the nominal frequency. 

\subsection{Filter by quality indicator}
The quality indicator should be used to filter the time series according to intended application. 
For statistical analyses it might be appropriate to eliminate all interpolated and non-valid entries ($\text{QI}\neq2$). 
In MATLAB this operation (including import) could be performed writing:
\begin{lstlisting}[]
Data = readtimetable('GB01.csv');
Data.f50_GB(Data.QI_GB ~= 0) = nan;
Data = rmmissing(Data);
\end{lstlisting}
For observing the trajectories instead, the build-in interpolation might be of help.
However, some campaigns of the data set are fragmented, so it is advisable to always plot the quality indicator if continuity of the frequency trajectory matters. 
We want to point out that the chosen linear interpolation should not yield a perfect imputation of missing values---we simply intend to give users an easy start in exploring this data set.

\section*{Acknowledgments}
We would like to express our gratitude to everyone who helped to create the data base by connecting the EDR in their hotel room, home, or office: 
Damià Gomila, 
Malte Schröder, 
Jan Wohland, 
Cigdem Yalcin, 
Filipe Pereira, 
André Frazão, 
Kaur Tuttelberg, 
Jako Kilter,  
Hauke Hähne, 
and Bálint Hartmann. 
We gratefully acknowledge support from the Helmholtz Association by the Program-Oriented Funding program ``Storage and Cross-Linked Infrastructures'', via the joint initiative ``Energy System 2050 - A Contribution of the Research Field Energy'', 
with the grant no. VH-NG-1025, and from the associative ``Uncertainty Quantification – From Data to Reliable Knowledge (UQ)'' with grant no. ZT-I-0029.
This work was performed as part of the Helmholtz School for Data Science in Life, Earth and Energy (HDS-LEE).
This project has received funding from the European Union's Horizon 2020 research and innovation program under the Marie Skłodowska-Curie grant agreement no. 840825.

\section*{Author contributions}
Conceptualization: RJ, HM, BS, and LRG. 
Development, construction of the EDR, and campaign logistics: HM and RJ. 
Conduction of measurement campaigns: BS, LRG, RJ, and HM. 
Data preparation and reprocessing: RJ and HM.
Technical validation: RJ.
Original draft preparation: RJ and BS. 
Review and editing: HM, BS, LRG, and VH. 
Supervision: HM and VH. 
Funding acquisition: VH. 
All authors have read and agreed to this version of the manuscript. 

\section*{Competing interests}
The authors declare no competing interests.



\end{document}